\documentclass[12pt,a4paper]{article}
\usepackage[utf8]{inputenc}
\usepackage[english]{babel}
\usepackage{amsmath}
\usepackage{amsfonts}
\usepackage{amssymb}
\usepackage{graphicx}
\usepackage{float}
\usepackage[left=3cm,right=2.5cm,top=2.5cm,bottom=2.5cm]{geometry}
\usepackage{epsfig}

\newcommand{\et }{{\(\eta\) }}

\newcommand{\urq }{{\footnotesize URQMD }}

\newcommand{\qgp }{{\footnotesize QGP }}

\newcommand{\qcd }{{\footnotesize QCD }}

\newcommand{\md}{{\footnotesize MD }}

\newcommand{\isr}{{\footnotesize ISR }}
\newcommand{\sps}{{\footnotesize SPS }}
\newcommand{\ags}{{\footnotesize AGS }}
\newcommand{\lhc}{{\footnotesize LHC }}
\newcommand{\ac}{{\footnotesize AA }}
\newcommand{\kno}{{\footnotesize KNO }}
\newcommand{\nbd}{{\footnotesize NBD }}
\newcommand{\rhic}{{\footnotesize RHIC }}
\newcommand{\hij }{{\footnotesize HIJING }}

\begin{document}

\begin{center}
\textbf{\large Entropy and multifractal spectrum in ring-like and jet-like events produced in $^{197}$Au-AgBr collisions at 11.6A GeV/c}\\
\vspace{5mm}
{\bf Bushra Ali, Sweta Singh, Anuj Chandra and Shakeel Ahmad\footnote{Shakeel.Ahmad@cern.ch}} \\
\vspace{3mm}
{\small Department of Physics, Aligarh Muslim University\\
Aligarh 202002, India}

\end{center}

\vspace{1.2cm}
{\footnotesize \noindent \textbf{Abstract:} Physical quantities such as, entropy, dimensions and multifractal characteristics of multiplicity distributions of charged particles produced in $^{197}$Au-AgBr collisions are examined and the findings are compared with the predictions of Monte Carlo model \urq (Ultra-Relativistic Quantum Molecular Dynamics) and \hij (Heavy Ion Jet INteraction Generator) and also with the results reported earlier in hadron-hadron and nucleus-nucleus collisions at different energies. Based on their azimuth distribution, the charged particles produced within narrow-bins exhibit two kinds of substructures, namely, ring-like and jet-like substructures. Thus,  on applying the suitable criteria the two different types of events are identified and analyzed separately. It is observed that the maximum entropy production occurs around a narrow mid-rapidity region. The analyses of ring-like and jet-like events suggest that the entropy production is much larger in ring-like events as compared to that in jet-like events. Furthermore, Renyi's order-q information entropy is used to estimate the multifractal specific heat and to construct the spectrum of scaling indices. The findings reveal that the value of multifractal specific heat is higher in ring-like events as compared to that in jet-like events. The studies of generalized dimension and multifractal spectrum indicate that the multifractality is rather, more pronounced in ring-like events as compared to jet-like events. Various features of the experimental data are noticed to be nicely reproduced by the \urq model.\\

\vspace{0.2cm}
}
\vspace{1mm}

\section*{Introduction}
\label{sec1}

\noindent The simplest and day-one observable which is readily accessible is the multiplicity of the relativistic charged particles produced in high energy hadronic (hh) and ion-ion ({\footnotesize AA}) collisions\cite{bib1,bib2}. By studying the multiplicity distributions ({\footnotesize MD}) of relativistic charged particles  for a given data sample, information on soft \qcd (Quantum Chromodynamics) processes as well as on hard scattering can be extracted\cite{bib2,bib3,bib4,bib5}.  Although numerous attempts have been made during the last few decades to study the features of \md of relativistic charged particles produced in hh and \ac collisions\cite{bib6,bib7,bib8,bib9,bib10,bib11,bib12,bib13,bib14,bib15,bib16}, yet a complete understanding of particle production mechanism still remains elusive. Asymptotic scaling of \md in hh collisions, referred to as the \kno scaling\cite{bib17} was regarded as a useful phenomenological framework to predict and compare \md in the incident energy range $\sim$ 10 GeV to \isr energies\cite{bib15,bib16}. After the observation of \kno scaling violation in $\rm\bar{p}p$ collisions at \sps energies\cite{bib18,bib19,bib20}, it was remarked that the observed scaling of \md up to \isr energies was approximate and accidental\cite{bib18}. To predict the \md at various energies, a new empirical regularity in place of \kno scaling was then proposed\cite{bib20}. It was observed that\cite{bib21,bib22} \md in limited and full pseudorapidity (\et) windows may be nicely reproduced by negative binomial distribution ({\footnotesize NBD}).  It has been observed by {\footnotesize CMS} collaboration\cite{bib23} that for pp collisions at \lhc energies $\sim$ (0.9 -- 13) TeV \kno scaling holds for small \et windows, whereas for large \et windows strong violations of \kno scaling are observed. The observed scaling violations are attributed to semi-hard gluon radiations (minijets) and multiparton scattering\cite{bib12}. Furthermore, it has been observed\cite{bib12} that at 0.9 TeV for $|\eta| <$ 1.3, \md for non single diffractive (nsd) events are not well fitted by a single \nbd and rather a parameterization with a sum of two \nbd be performed  for the nsd, inelastic (inel) and inel $> 0$ event samples.  
{\footnotesize ALICE} collaboration\cite{bib12}, while studying the energy dependence of \(C_q\) moments at 0.2, 0.9 and 2.36 TeV, has reported that for non single diffractive events  {\footnotesize KNO} scaling gives a reasonable description of the data from 0.2 to 2.36 TeV. However, the ratio \(P(z)\langle N_{ch}\rangle\) between 0.9 TeV and 2.36 TeV data has been reported to show slight departure from unity above \(z = 4\); here \(z(= N_{ch}/\langle N_{ch}\rangle)\) is the {\footnotesize KNO} variable. The collaboration, in its detailed study\cite{bib13}, has observed that for nsd events, {\footnotesize KNO} scaling violation increases with increasing  \(\eta\) intervals. The shape of {\footnotesize KNO} scaling violation reflects the fact that the high multiplicity tail of the distribution increases faster with increasing energy and \(\eta\) interval than the low (\(N_{ch} \le 10\)) multiplicity part\cite{bib13}. It has also been observed by {\footnotesize ALICE} collaboration\cite{bib24} that  {\footnotesize KNO} scaling is violated for all pseudorapidity intervals for both soft and semihard components in the nsd, inel and inel \(>\) 0 event samples. However, analysing the \(pp\) data from {\footnotesize ATLAS} at \lhc, Kulchitsky and Tsiareshka\cite{bib25} have observed that for \(|\eta| < 2.5\) and \(z > 1\), the  {\footnotesize KNO} scaling is valid. These observations, thus, led to a revival of interest in investigations involving \md and new scaling laws. \\

\noindent Simak et al\cite{bib26}, by introducing a new variable --the information entropy, showed that \md of charge particles produced in full and limited \et ranges in hh collisions exhibit a new type of scaling in the range $\sqrt{s} \sim$ (19 -- 900) GeV. Sinyukov and Akkelin\cite{bib27} introduced a method to evaluate the entropy of thermal pions in \ac collisions. They examined the average phase space densities and entropies of such pions against their multiplicities and beam energies.  The findings indicate the presence of  deconfinement and chiral phase transition in \ac collisions at relativistic energies. Furthermore, at \rhic energies, entropy per unit rapidity at freeze out has been extracted with minimal model dependence from the available measurements of particle spectra yields and source sizes, determined by two-particle interferometry\cite{bib28}. The estimated entropy per unit rapidity was found to be consistent with the Lattice Gauge Theory for thermalised \qgp with the energy density calculated using the transverse energy production at \rhic  energies. \\

\noindent Investigation involving entropy production in pp, $\rm\bar{p}$p and $\pi^{\pm}$p/k$^{\pm}$p collisions over a wide range of  beam energies (up to $\sqrt{s} =$ 900 GeV)\cite{bib7,bib9,bib10,bib26} indicates that the entropy produced in full and limited phase spaces increases with incident energy, whereas the entropy per unit rapidity appears to be an energy independent quantity. These findings, thus, tend to suggest  the entropy scaling up to a few TeV of energy. Mizoguchi and Biyajima and Das et al\cite{bib7} have also observed a similar entropy scaling in pp collisions at \lhc energies. \\

\noindent As for the studies involving \ac collisions are concerned, the main aim is to study the properties of strongly interacting matter under extreme conditions of nuclear density and temperature, where \qgp is expected to be formed\cite{bib1,bib6,bib11,bib29,bib30,bib31,bib32}. Fluctuations in physical observables in \ac collisions are regarded as one of the important signals for \qgp formation because of the idea that in many body systems, phase transition results in significant changes in quantum fluctuations of an observable from its average behavior\cite{bib2,bib11,bib29,bib33}. For example, when a system undergoes a phase transition, heat capacity changes abruptly, while the energy density remains a smooth function of temperature\cite{bib6,bib11,bib28,bib34,bib35}. Entropy is regarded as yet another important characteristic of the system with many degrees of freedom\cite{bib6,bib36,bib37,bib38}. Systems, in which particles are generally produced, may be regarded as the so-called dynamical systems\cite{bib2,bib36,bib37,bib38,bib39}  in which entropy is generally produced. Systematic measurements of local entropy produced in \ac collisions may provide direct information about the internal degrees of freedom of the \qgp medium and its evolution\cite{bib2,bib6,bib11,bib40}.  Particles produced in high energy collisions have been argued\cite{bib41} to occur at maximum stochasticity, i.e., they follow the maximum entropy principle.  This type of stochasticity may also be  quantified in terms of information entropy which may be regarded as a natural and more general parameter to measure the chaoticity in branching processes\cite{bib42}.\\

\noindent It has also been proposed\cite{bib3,bib8,bib37,bib43} that Renyi's order-q entropies may also serve as a good tool for studying dynamical systems and are closely related to thermodynamic entropy  -- the Shannon entropy of the system. Furthermore, the generalization of Renyi's order-q information entropy contains information on the multiplicity moments that may be used to examine the multifractal characteristics of multiparticle production\cite{bib44,bib45,bib46,bib47}. It should be mentioned here that this method of study of the multifractality is not related to the phase space bin width and (or) the detector resolution, whereas in conventional intermittency studies using the scaled factorial movements, $F_q$, a linear dependence on the bin resolution ($\delta$) of the form $ln F_q =  - \tau_{q} ln\ \delta$ is looked into. Such behavior is interpreted as (multi)fractal property of particle production\cite{bib10}.\\

\begin{figure}[th]%
\centering
\centerline{\psfig{file=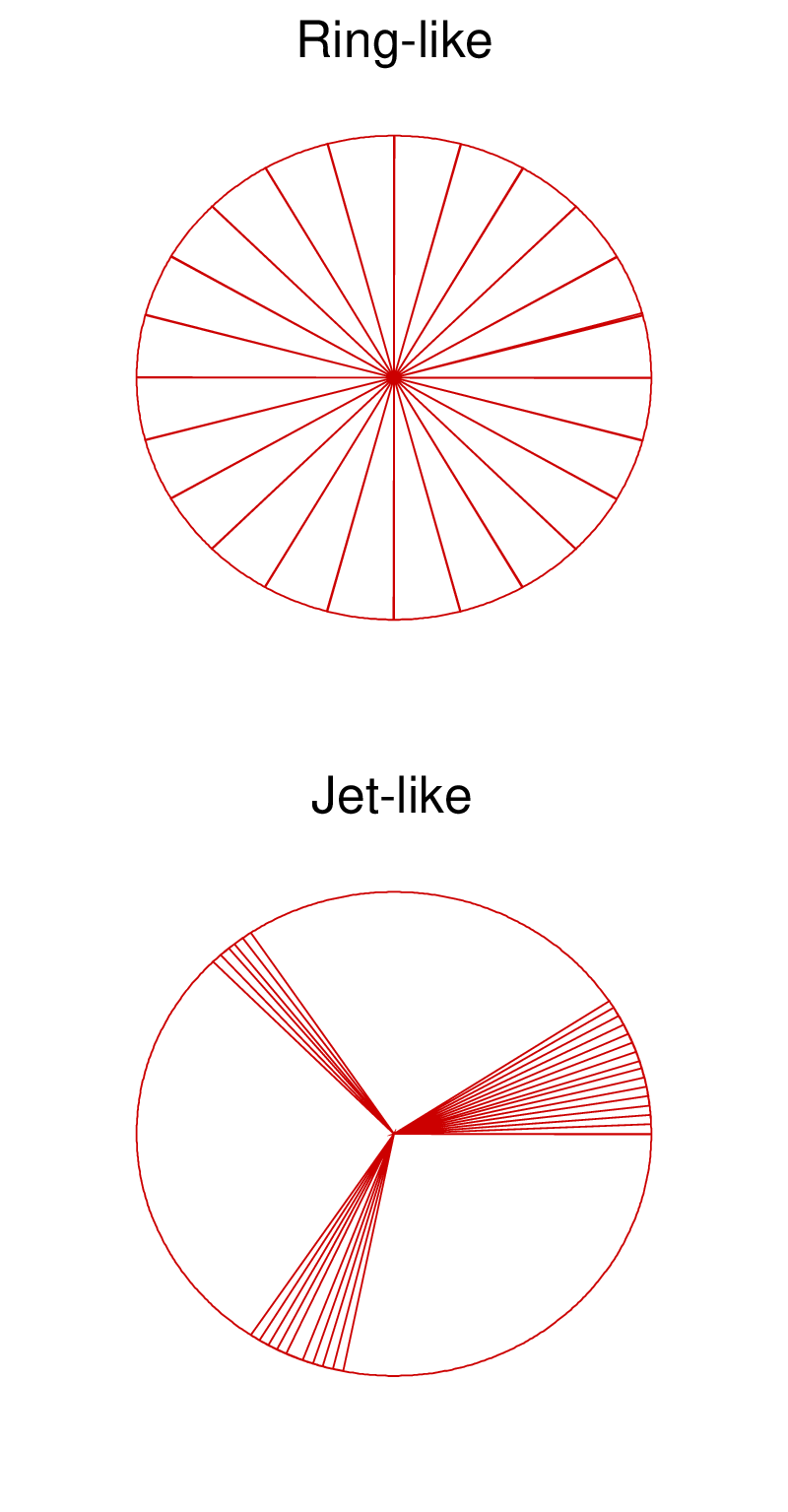,height=10cm}}
\caption{Schematic diagram of ring-like and jet-like events.}\label{fig1}
\end{figure}

\noindent Entropy production in \ac collisions at \ags and \sps energies has  been examined earlier\cite{bib1,bib2,bib11,bib33}. These findings reveal that the entropy in limited and full phase spaces, if normalized to maximum rapidity exhibits a kind of scaling. It has also been shown that the Renyi's order-q information entropy could be another way to investigate the \md in terms of multifractal spectrum $f(\alpha_q)$, which has been argued to be a convenient function for comparing not only the results from different experiments but also between the data and the theoretical models. It was, therefore, considered worthwhile to carry out a detailed study of entropy production and multifractal  characteristic of multiparticle production in $^{197}$Au-AgBr collisions at 11.6A GeV/c and comparing the findings with the predictions of Monte Carlo models \urq and \hij.\\

\noindent As the azimuth distribution of pions within a narrow \et bin exhibits two different substructures, namely, the ring-like and the jet-like\cite{bib48,bib49}. The events having the two types of substructures are sorted out and analyzed separately. If many particles are produced within a narrow \et bin but spread over the entire azimuth,  a ring-like substructure will appear. However, if particles within a narrow \et window are also confined to a narrow $\phi$ region, a jet-like substructure would emerge\cite{bib31,bib48,bib49,bib50,bib51,bib52}. A schematic diagram of ring-like and jet-like substructures is presented in Fig.~\ref{fig1}. Ring-like substructures are envisaged to occur due to Cerenkov gluons. Each gluon gives rise to a jet. These jets are expected to create a ring with $``$jetty$"$ spots or $``$jetty$"$ substructures in the azimuth plane which is perpendicular to the primary parton orientation. A large number of Cerenkov gluons are produced at very high energy and may form a ring in a single event. Coherent collective effects in hadronic matter may also give rise to the so-called ring-like events\cite{bib31,bib48,bib51,bib52,bib53,bib54,bib55}.  Although the ring-like and jet-like events do not show noticeable deviations as is expected from their isotropic nature. It may be emphasized here that investigations involving both types of substructures do not give complete information about the processes involved and hence the two types of events should be analyzed separately\cite{bib51}.\\

\noindent The ring-like and jet-like events may be identified from a given data sample by following the method proposed by Adamovich et al\cite{bib49}. According to this approach, a fixed number of particles, $n_d$ are considered. This $n_d$ tuple of particles along the \et axis is taken as a group characterized by $\Delta\eta_c$. Thus, the particle density in this \et interval is $\rho_c = n_d/\Delta\eta_c$. Since the multiplicity of particles in each subgroup does not depend on density, therefore, it can be compared with each other for a given subgroup. For a given subgroup, the azimuth structure is to be parameterized in such a way that the larger values of parameter refer to one type of substructure, whereas the smaller values refer to the other type. The following two sums have been proposed\cite{bib56} for these parameters: 

\begin{eqnarray}
 S_1 = -\sum ln(\Delta\Phi_i)
\end{eqnarray}
and
\begin{eqnarray}
 S_2 = \sum ln(\Delta\Phi_i)^2
\end{eqnarray}

\noindent where $\Delta\Phi_i$ denotes the azimuth difference between two neighboring particles in the group. For simplicity sake, $\Delta\Phi_i$ is counted in units of full revolutions, which would yield :
\begin{eqnarray}
 \sum(\Delta\Phi_i) = 1
\end{eqnarray}

\noindent Both the parameters $S_1$ and $S_2$ will be small ($S_1 = \rm{n_d} ln\ \rm{n_d}, S_2 \rightarrow 1/n_d$) for ring-like substructure and large ($S_1 \rightarrow \infty, S_2 \rightarrow 1$) for a jet-like structure. Although $S_1$ and $S_2$ have similar properties, $S_1$ is sensitive to the smallest gap, $\Delta\Phi_i$, whereas the major contribution to $S_2$ comes from the largest gap or void in the group\cite{bib49}. The ring-like and jet-like events are sorted out by calculating the value of $S_2/\langle S_2\rangle$ on an event-by-event (ebe) basis. The events with $S_2/\langle S_2\rangle > 1$ are taken as jet-like events while the event with $S_2/\langle S_2\rangle < 1$ are categorized as ring-like events.\\

\section{Details of the data}
\noindent A sample consisting of 577 events produced in the interactions of 11.6A GeV/c $^{197}$Au beam with the AgBr group of nuclei in nuclear emulsion are used for the present study. This data set is taken from the series of experiments carried out by the {\footnotesize EMU01} collaboration\cite{bib57,bib58,bib59,bib60}. All the relevant details, like criteria for selection of events, track classification, extraction of AgBr group of events, method of measurements, etc., may be found elsewhere\cite{bib1,bib2,bib31,bib32,bib61}. It may be emphasized here that the conventional emulsion technique has two main advantages over the other detectors: $i)$  its $4\pi$ solid angle average and $ii)$ the data is free from biases due to full phase space coverage. In the case of other detectors, only a fraction of charged particles are captured due to their limited acceptance cone. This not only reduces the charge particle multiplicity but may also distort some of the event characteristics, like particle density fluctuations\cite{bib1,bib62}. For comparing the findings of the present work with the  predictions of the theoretical models, {\footnotesize URQMD}\cite{bib63,bib64} and {\footnotesize HIJING}\cite{bib65,bib66}, Monte Carlo (MC) event sample corresponding to real data is simulated using the code \urq-3.4 and \hij-1.35. The number of events in the simulated sample is kept the same as that in the experimental data. The events are simulated by taking into account the percentage of interactions occurring with various  target nuclei in emulsion\cite{bib67,bib68}. The values of the impact parameter are so set, while generating the MC data, that the mean multiplicity of relativistic particles nearly match with those obtained from the real data. Ring-like and jet-like events are separated by following the criteria discussed in section-1. The proportions (in $\%$) of the two categories of events, ring-like and jet-like are found to be 60:40, 58:42 and 60:40 respectively for real, \urq and \hij event samples.\\

\section{Formalism}
\noindent Values of Shannon's information entropy are evaluated using the relation\cite{bib1,bib2,bib10,bib26}: \\
\begin{eqnarray}
S = -\Sigma P_n ln P_n
\label{eq4}
\end{eqnarray}

\noindent and its generalization, Renyi's order-q information entropy is calculated as\cite{bib2,bib9,bib10}:\\
\begin{eqnarray}
I_q = \frac{1}{1-q} ln \Sigma_n P_n^q
\label{eq5}
\end{eqnarray}

\noindent where, for q = 1, $lim_{q \rightarrow 1} I_q = I_1 = S$ , while $P_n$ is the probability of production of charged particles in a given \et window. The generalized dimension of order q is then estimated as:\\
\begin{eqnarray}
D_q = \frac{I_q}{Y_m}
\label{eq6}
\end{eqnarray}
where,
\begin{eqnarray}
Y_m = ln \left[\frac{(\sqrt{s} - 2 m_n\langle n_p\rangle)}{m_{\pi}}\right] = ln(n_{max})
\label{eq7}
\end{eqnarray}
$Y_m$ is referred to as the maximum rapidity in the center-of-mass frame, $\sqrt{s}$ represents center-of-mass energy, $m_n$ and $m_\pi$ are the masses of nucleons and pion respectively, $\langle n_p\rangle$ denotes the mean of the number of participating nucleons and $n_{max}$ is the maximum multiplicity of the relativistic charged particles produced. It is evident from Eq.\ref{eq5} that for a given q, $(I_q)_{max} = ln\ n_{max}$. The maximum entropy is achieved for the greatest $``$chaos$"$ of a uniformly distributed probability function $P_n = 1/n_{max}$. Thus, Eq.\ref{eq6} gives $D_q = I_q/ (I_q)_{max}$\cite{bib46,bib47}. \\

\noindent If the particle production process follows self-similar behavior, the multifractal moments of order q, defined as:\\

\begin{eqnarray}
G_q = \Sigma P_n^q 	\quad\quad \rm ; q\ is\ any\ real\ number
\label{eq8}
\end{eqnarray}

\noindent should exhibit the following power law behavior
\begin{eqnarray}
G_q \propto (\delta \eta)^{\tau(q)}
\label{eq9}
\end{eqnarray}

\noindent where $\delta\eta$ is \et bin width while the parameter $\tau_q$ is related to the dimension $D_q$ for all q by:

\begin{eqnarray}
\tau_q(q) = (q-1) D_q
\label{eq10}
\end{eqnarray}

\noindent where $D_0$, $D_1$ and $D_2$ are usually referred to as the fractal dimension, information dimension and correlation dimension respectively\cite{bib69}.\\

\noindent It has been pointed out by Hwa\cite{bib69} that the values of $G_q$ moments obtained from various experiments can not be compared as they depend on the number of events in the data and on $\delta\eta$, i.e., on detector resolution. The aim of studying the $G_q$ dependence on $\delta\eta$ is to estimate the generalized dimensions, $D_q$, where
\begin{eqnarray}
D_q = (q-1) lim_{\delta\eta \rightarrow 0} \left(\frac{ln\ G_q}{ln\ \delta\eta}\right)
\label{eq11}
\end{eqnarray}

\noindent The meaning of the function $\tau(q)$ becomes rather more obvious after performing the Legendre transformation from independent variables $\tau$ and $q$ to the variables $\alpha$ and $f$ as:

\begin{eqnarray}
\alpha_q = \frac{d\tau(q)}{dq} \\
f(\alpha_q) = q\alpha_q - \tau(q)
\label{eq12}
\end{eqnarray}

\noindent where $f(\alpha_q)$ is the fractal dimension of a subset composed from bins whose occupancy probability lies in the range, $(P - dP)$ to $(P + dP)$. Once the values of $G_q$ moments are obtained, continuous scaling function $f(\alpha_q)$ can be constructed\cite{bib2,bib70}. The thermodynamical interpretation of these relations implies that q can be related to inverse temperature, $q = T^{-1}$, whereas the spectrum $f(\alpha_q)$ and $\alpha$ play the role of entropy and energy (per unit volume), respectively\cite{bib71,bib72,bib73,bib74}. \\

\begin{figure}[b!]
\centerline{\psfig{file=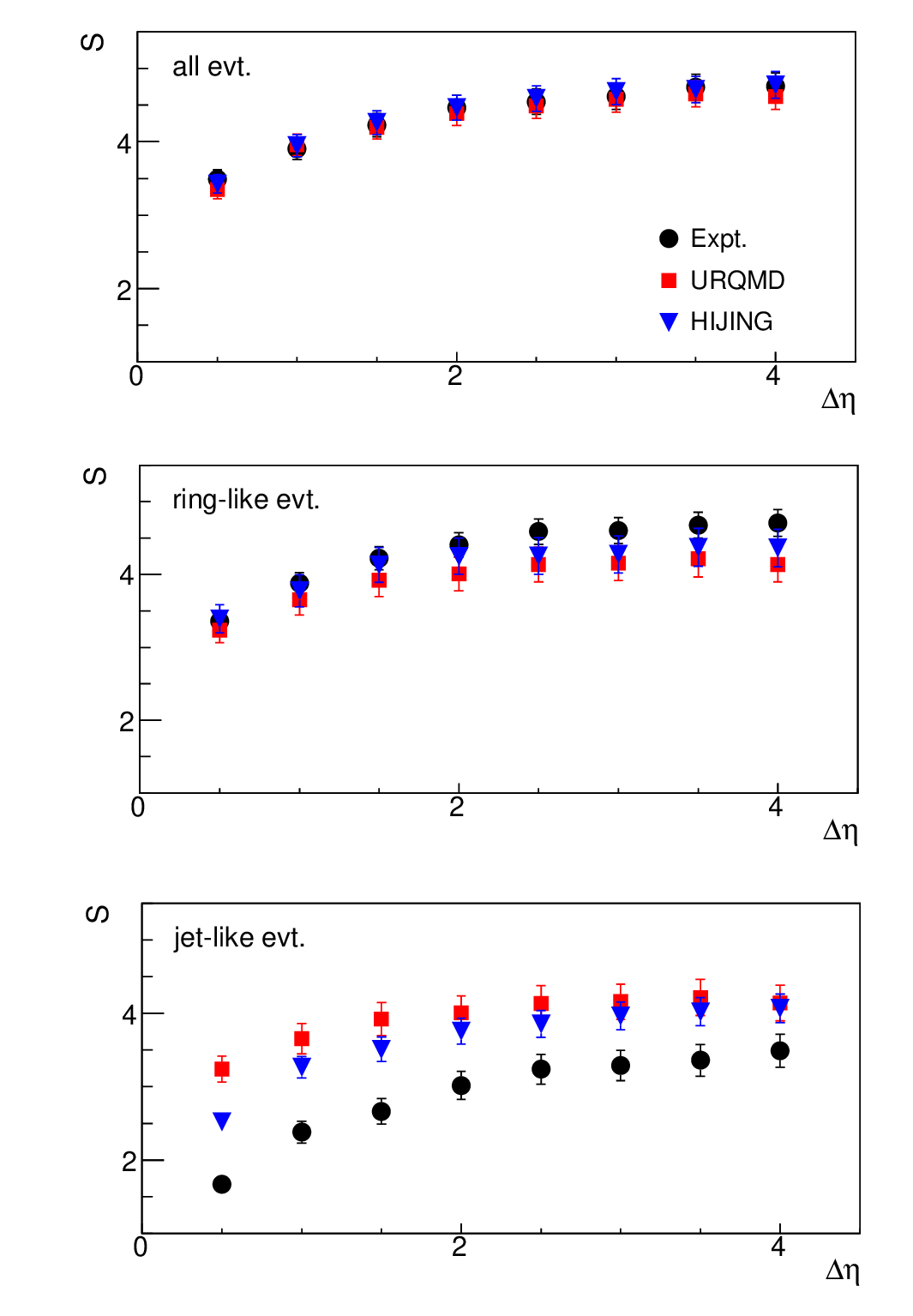,height=14cm}}
\caption{Variations of S with $\Delta\eta$ for the experimental, URQMD and HIJING. Shown in the middle and bottom panels are the variations for ring-like and jet-like events.}\label{fig2}
\end{figure}
\section{Results and Discussions}
\noindent MD of charged particles produced in a \et window of fixed width $\Delta\eta=0.5$ is obtained. This window is so selected that its mid-position coincides with the center of symmetry of \et distribution. Thus, all the charged particles having their \et values in the range, $(\eta_c - \frac{\Delta\eta}{2}) < \eta < (\eta_c + \frac{\Delta\eta}{2})$ are counted to calculate $P_n$, which in turn is used to evaluate $S$ using Eq.\ref{eq4}. The window width is then increased in steps of 0.5 \et units, until the region $\eta_c \pm 2.0$ is covered. Variations of $S$ with $\Delta\eta$ for ring-like, jet-like and all events are displayed in Fig.\ref{fig2}. Data points shown in the top panel correspond to real data, while the ones lying the middle and bottom panel of the figure are due to \urq and \hij event samples. It may be noted that $S$ grows with $\Delta\eta$ upto $\Delta\eta \sim 2$ and then tends to acquire a saturation: it is interesting to note that the models, \urq and \hij nearly reproduce the trend exhibited by the real data. Moreover, for any given $\Delta\eta$, values of $S$ for ring-like events are found to be slightly higher as compared to those obtained for all events. Jet-like events are however, noticed to acquire much smaller values of $S$ as compared to those estimated for all events. It has been observed that the data used in the present study and the data sets involving $^{16}$O-AgBr and $^{28}$Si-AgBr collisions at 14.5A GeV/c,  $^{16}$O-AgBr collisions at 60A GeV/c and $^{16}$O-AgBr and $^{32}$S-AgBr collisions at 200 GeV/c, values normalized to maximum rapidity, when plotted against $\Delta\eta$ also normalized to maximum rapidity exhibit a kind of entropy scaling in \ac collisions at \ags and \sps energies\cite{bib1,bib2,bib11}. Similar entropy scaling in \ac collisions at these energies has also been reported by other workers too for central and minimum bias events\cite{bib33,bib71}.\\

\begin{figure}[b!]
\centerline{\psfig{file=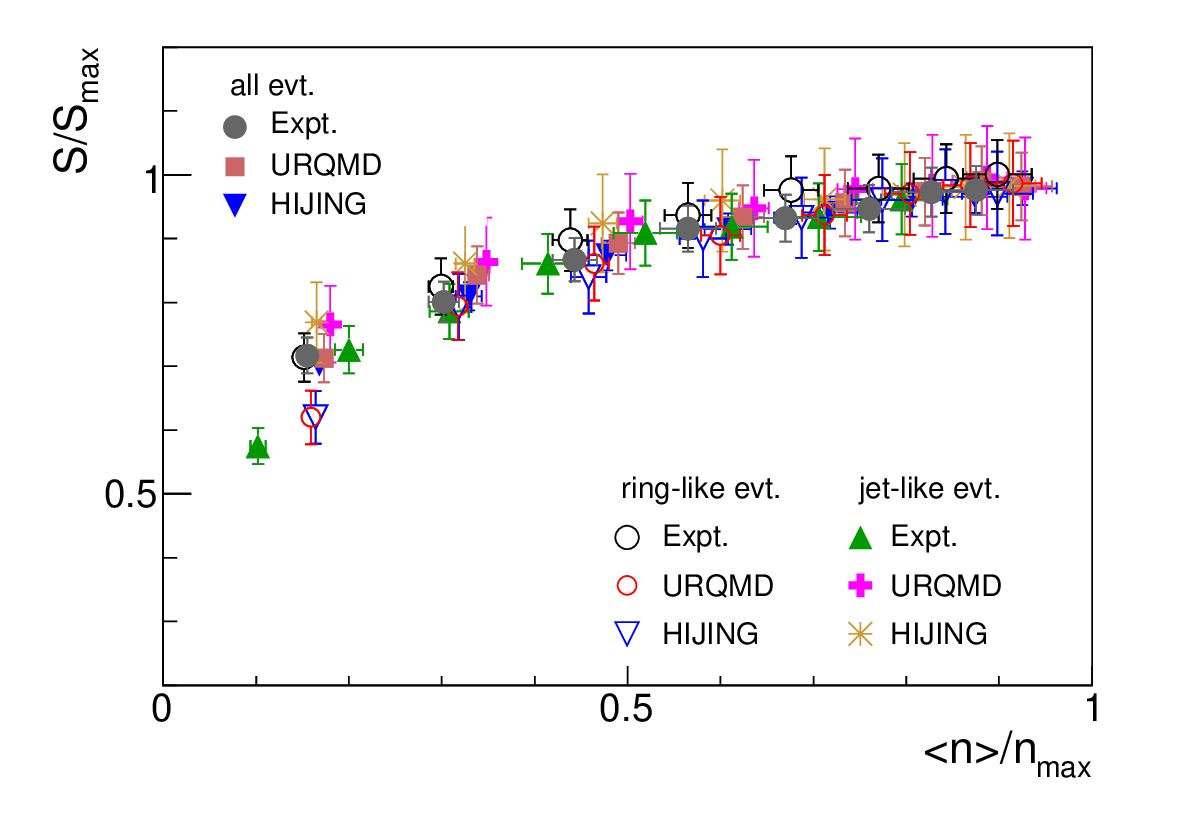,height=8.0cm}}
\caption{Variations of $S/S_{max}$ with $\langle n\rangle/n_{max}$ for the real, URQMD and HIJING data and also for the ring-like and jet-like events.}\label{fig3}
\end{figure}

\noindent According to Eq.\ref{eq7}, the quantity ($\sqrt{s} - 2m_n\langle n_p\rangle/m_{\pi}$) is equal to the maximum charged particle multiplicity for a given data set. It would be convenient to examine the entropy dependence on mean multiplicity in limited and full \et ranges. The entropy in full \et range, $S_{max}$ is calculated using Eq.\ref{eq4}. Variations of $\frac{S}{S_{max}}$ with $\frac{\langle n\rangle}{n_{max}}$ for the real, \hij and \urq data and also for the ring-like and jet-like events are plotted in Fig.\ref{fig3}. It is evident from the figure that data points corresponding to various sets of events overlap to form a single curve. Furthermore, it was noticed that $\frac{S}{S_{max}} \rightarrow 1$ as $\frac{\langle n\rangle}{n_{max}} \rightarrow 1$. These observations, therefore, tend to support the presence of entropy scaling in \ac collisions at \ags and \sps energies.\\

\begin{figure}[b!]
\centerline{\psfig{file=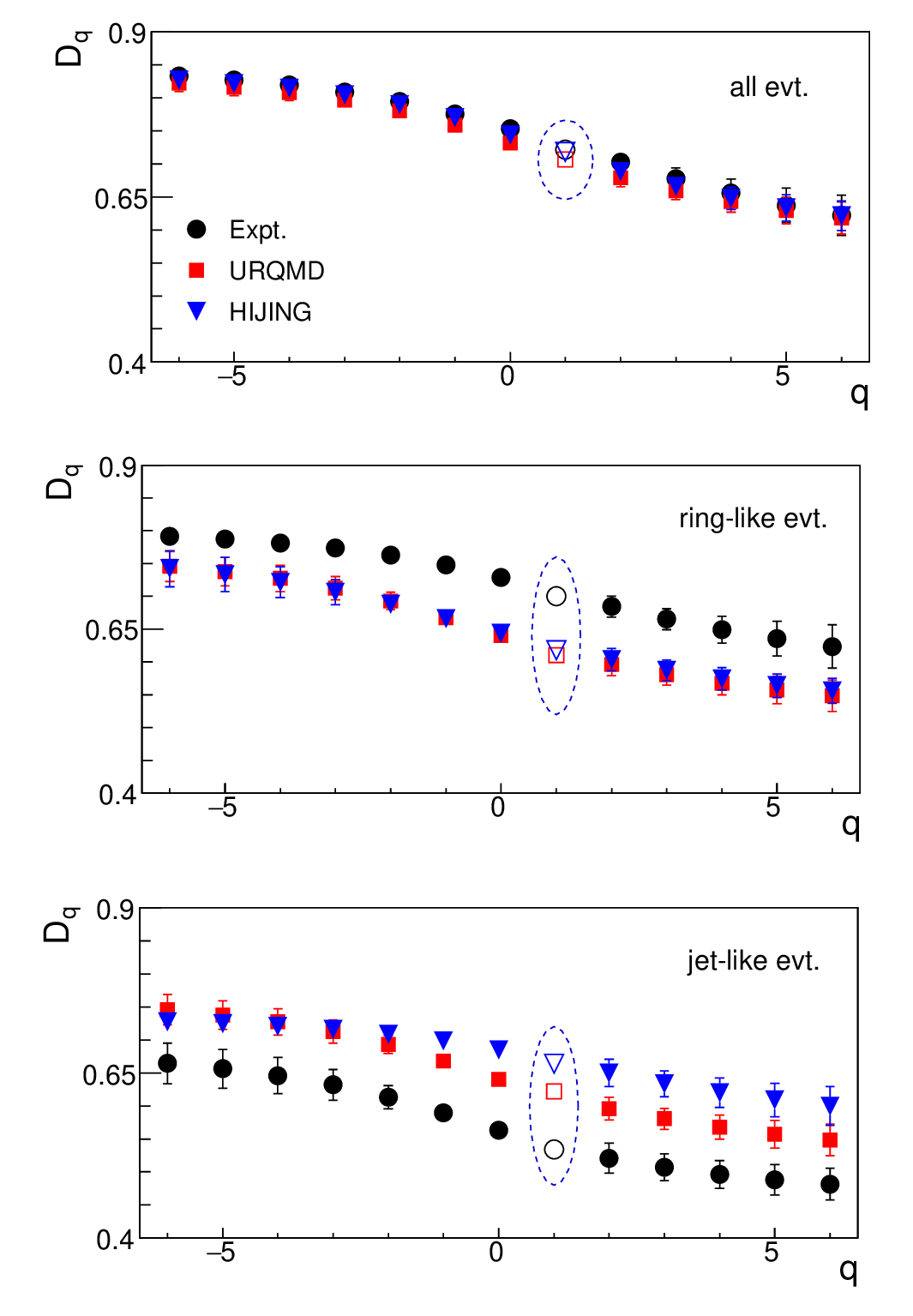,height=16cm}}
\caption{Dependence of $D_q$ on q for various data sets. The encircled open markers represent the values of $D_1$ obtained from $f(\alpha_q)$ spectra.}\label{fig4}
\end{figure}

\noindent Values of $D_q$ for -6 $\leqslant q \leqslant$ 6 are calculated using Eqs. 5-7. Variations of $D_q$ with q for various data sets are shown in Fig.\ref{fig4}. The values of $D_1$ have been obtained from the $f(\alpha_q)$ spectra, as described in coming part of the text. It is noticed from the figure that the values of $D_q$ monotonically decrease with increasing order q. It may also be noted that model predicted values are quite close to those estimated using real data. It may also be observed that although the trends of variations of $D_q$ with $q$ for ring-like and jet-like events are in qualitative agreement with those exhibited by all events. However, in comparison to real data, model predicted values of $D_q$ for a given $q$ are larger for jet-like events but somewhat smaller for ring-like events. Moreover, the $D_q$ spectra for $q \geq 2$, which for multifractals is a decreasing function of q, may be related to the scaling behavior of q point correlation integrals\cite{bib10,bib75}. The observed trends of $D_q$ dependence on q observed in the present study, therefore, tend to suggest the multifractal nature of \md. Similar trends of variations of $D_q$ with q have also been reported for pp (800 GeV) and heavy-ion collisions in a wide range of incident energies\cite{bib46,bib47}. For a given \(\Delta\eta\), nearly similar values of \(D_q\) and \(S\) for all and ring-like events might be due to similar shapes of multiplicity distributions of the two types of events within the considered \(\eta\) range. Although the decreasing trend of $D_q$ with q indicates the presence of multifractality, yet no further useful inference from the $D_q$  dependence on q can be arrived at, which may lead to some meaningful conclusions on the scaling properties of q-correlation integrals\cite{bib2,bib10}.\\

\begin{table}[t!]
\centering
\caption{The values of parameters, a and c, occurring in Eq.\ref{eq14}}\label{tab1}%
\begin{tabular}{@{}rlcc@{}}
\hline
\multicolumn{2}{c}{Event Type}  &  a   & c \\
\hline
            & Expt. & 0.789 $\pm$ 0.019 & 0.261 $\pm$ 0.018 \\
   All Evt. & \urq  & 0.755 $\pm$ 0.014 & 0.198 $\pm$ 0.013 \\
            & \hij  & 0.768 $\pm$ 0.012 & 0.212 $\pm$ 0.011 \\
\hline
            & Expt. & 0.752 $\pm$ 0.017 & 0.201 $\pm$ 0.015 \\
  Ring-like & \urq  & 0.647 $\pm$ 0.011 & 0.155 $\pm$ 0.010 \\
            & \hij  & 0.652 $\pm$ 0.007 & 0.149 $\pm$ 0.006 \\
\hline
            & Expt. & 0.560 $\pm$ 0.006 & 0.123 $\pm$ 0.005 \\
   Jet-like & \urq  & 0.688 $\pm$ 0.011 & 0.195 $\pm$ 0.010 \\
            & \hij  & 0.704 $\pm$ 0.011 & 0.162 $\pm$ 0.010 \\
\hline
\end{tabular}
\end{table}

\begin{figure}[th]
\centerline{\psfig{file=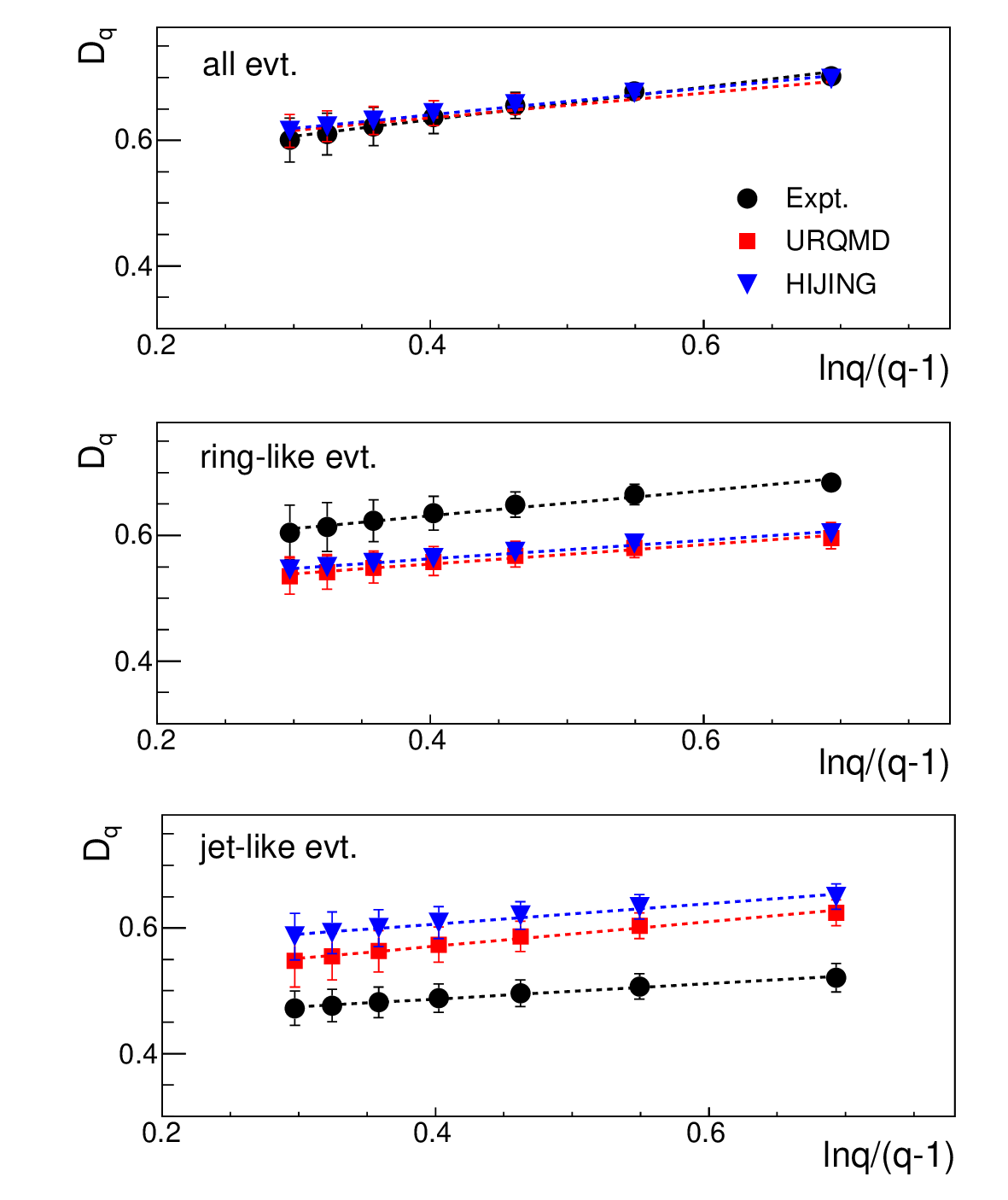,height=15cm}}
\caption{Variations of $D_q$ with $ln\ q/(q-1)$ for the experimental, MC, ring-like and jet-like events. The lines represent the best fits to the data obtained using Eq.\ref{eq14}}\label{fig5}
\end{figure}

\noindent It has been suggested\cite{bib71,bib76,bib77} that in constant specific heat (CHS) approximation, $D_q$ dependence on q would acquire the following form: 
\begin{eqnarray}
D_q \simeq (a-c) +c\frac{ln\ q}{q-1}
\label{eq14}
\end{eqnarray}
\noindent where a is the information dimension $D_1$ while c is referred to as the multifractal specific heat. Such a linear trend of variation of $D_q$ with $ln\ q/(q-1)$ is expected to be observed for multifractals on the basis of classical analogy with specific heat of gases and solids. The values of c is predicted to be independent of temperature in a wide range of q\cite{bib78}. In order to test the validity of Eq.\ref{eq14}, variations of $D_q$ against $ln\ q/(q-1)$ are plotted for various data sets in Fig.\ref{fig5}. The lines in the figures are due to the best fits to the data, obtained using Eq.\ref{eq14}. The values of parameters, a and c, occurring in Eq.\ref{eq14} are presented in Table.\ref{tab1}. It is interesting to note that the value of multifractal specific heat obtained for the real data is somewhat higher as compared to ones predicted by \hij and \urq models. Furthermore, the value of c for the ring-like events is also close to that obtained for the entire data sample. The jet-like events, however, give somewhat smaller values of c. The value of c $\sim$ 0.2 obtained in present study are found to be close to those reported by Du et al\cite{bib77} for $^{197}$Au-nucleus collisions at the same beam energy and also for $^{16}$O-AgBr collisions at 14.5A, 60A and 200A GeV/c and $^{32}$S-AgBr collisions at 200A GeV/c. For p-nucleus collisions in the energy range $\sim$ (200 -- 800) GeV\cite{bib46,bib76,bib77}, the values of c have been reported to be $\sim$ 0.25. For pp/$\rm\bar{p}$p collisions, however, the values of this parameter have been observed to be $\sim$ 0.08 in the energy range 25-1800 GeV\cite{bib10}. These observations, therefore, tend to suggest that the constant specific heat approximation is applicable to the multiparticle production in relativistic hadronic and heavy-ion collisions. Moreover, the fact that nearly the same values of c are obtained for \ac collisions ($\sim$ 0.2) and pp collisions ($\sim$ 0.08) over a wide range of beam energies indicate that the parameter c may be taken as a universal characteristics of hadronic and heavy-ion collisions.\\

\begin{figure}[t!]
\centerline{\psfig{file=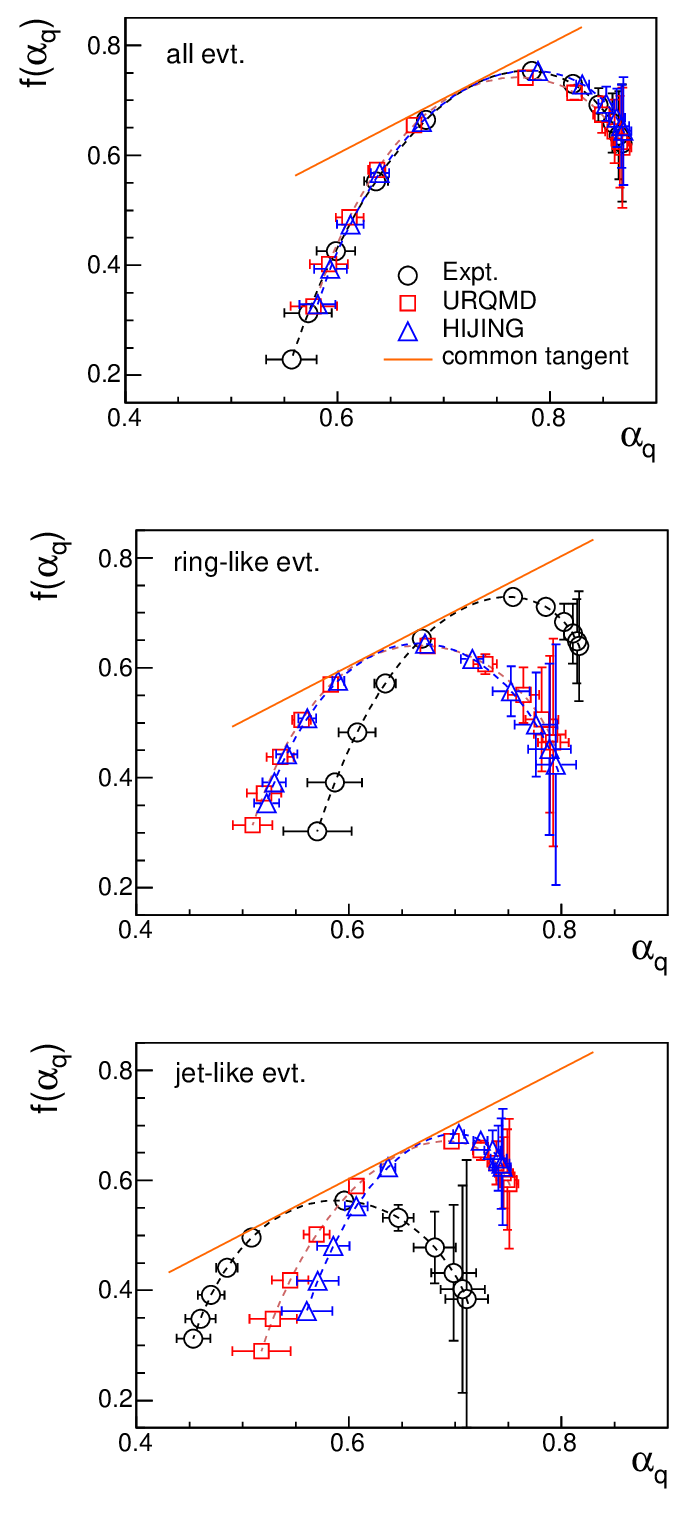,height=16cm, width=7cm}}
\caption{Multifractal spectra for the real, URQMD and HIJING events (top panel). Spectra due to jet-like events are displayed in the middle and bottom panels. The dotted lines represent the common tangent.}\label{fig6}
\end{figure}

\noindent Values of $\tau_q$, $\alpha_q$ and $f(\alpha_q)$ for -6 $\leqslant q \leqslant$ 6 are calculated using Eqs.10, 12 and 13 for constructing the multifractal spectrum. The spectra, thus, obtained are displayed in Fig.\ref{fig6}. It may be noted from the figure that the $f(\alpha_q)$ spectra for the real, \urq and \hij data nearly overlap. It is also evident from the figure that the spectrum for ring-like events is broader as compared to jet-like events. These spectra are seemed to form continuous curves, thus, characterizing a qualitative manifestation of multifractal fluctuations. These spectra are also noticed to follow the general characteristics of the occurrence of peaks at $\alpha_0$ and a common tangent at an angle 45$^{\circ}$, i.e., at $f(\alpha_1) = \alpha_1$. The spectra are also noticed to be concave downward everywhere with respect to the peak at $\alpha_0$. The region $\alpha_q < \alpha_0$ corresponds to positive q values and the curves in this region have positive slopes, while the region  $\alpha_q > \alpha_0$ corresponds to negative q values and the slope of the curves in this region is negative. The values of $f(\alpha_q)$ for q=0,1,2 respectively give the fractal dimension ($D_0 = f(\alpha_0)$), the information dimension ($D_1 = f(\alpha_1) = \alpha_1$) and the correlation dimension ($D_2 = 2\alpha_2 - f(\alpha_2)$). The values of information dimension, $D_1$ are shown in Fig.\ref{fig4} (encircled points). The width of $f(\alpha_q)$ spectra ($\alpha_{max} - \alpha_{min}$) for various events are calculated and listed in Table.\ref{tab2}. The width of the spectrum is taken as a measure of degree of multifractality\cite{bib79,bib80,bib81}. It is noted from the figure that the width of the spectra corresponding to \hij and \urq events are nearly equal to that obtained for the real data. It is also noted that the spectrum for ring-like events is relatively broader than that for jet-like events. This result suggests that the multifractality is more pronounced in ring-like events as compared to jet-like events.\\

\begin{table}[t!]
\centering
\caption{Values of \(\alpha_{min}\) and \(\alpha_{max}\) for various categories of events.}\label{tab2}%
\begin{tabular}{@{}rlccc@{}}
\hline
   \multicolumn{2}{c}{Event Type}  &  \(\alpha_{min}\)   & \(\alpha_{max}\) & $\left( \alpha_{max} - \alpha_{min} \right) $  \\
\hline
            & Expt.  & 0.557 $\pm$ 0.012 & 0.869 $\pm$ 0.004 & 0.312 $\pm$ 0.013 \\
   All Evt. & URQMD  & 0.578 $\pm$ 0.011 & 0.868 $\pm$ 0.004 & 0.290 $\pm$ 0.012 \\
            & HIJING & 0.581 $\pm$ 0.009 & 0.869 $\pm$ 0.004 & 0.288 $\pm$ 0.010 \\
\hline
            & Expt.  & 0.570 $\pm$ 0.016 & 0.817 $\pm$ 0.004 & 0.247 $\pm$ 0.016 \\
  Ring-like & URQMD  & 0.509 $\pm$ 0.009 & 0.793 $\pm$ 0.007 & 0.284 $\pm$ 0.011 \\
            & HIJING & 0.523 $\pm$ 0.006 & 0.795 $\pm$ 0.010 & 0.272 $\pm$ 0.012 \\
\hline
            & Expt.  & 0.454 $\pm$ 0.008 & 0.711 $\pm$ 0.010 & 0.257 $\pm$ 0.013 \\
   Jet-like & URQMD  & 0.518 $\pm$ 0.014 & 0.751 $\pm$ 0.004 & 0.233 $\pm$ 0.014 \\
            & HIJING & 0.560 $\pm$ 0.012 & 0.745 $\pm$ 0.004 & 0.185 $\pm$ 0.013 \\
    
\hline
\end{tabular}
\end{table}

\noindent Since $D_q$ values are observed to be larger for ring-like events, $\alpha_0$ and $f(\alpha_0)$ will also be higher for such events and hence the entire spectrum will be broader\cite{bib69}. This, in turn, suggests that the \et distribution for this category of events will be more jagged and irregular with deep valleys and sharp peaks. It has been suggested\cite{bib69} that the highly chaotic behavior of \md in narrow \et bins n(\et) may be observed by smooth function $f(\alpha)$, which is quite interesting. It has also been remarked\cite{bib2,bib69} that averaging n(\et) over all events would be devoid of any information related to fluctuations and intermittency and $f(\alpha_q)$ spectrum would appear somewhat narrower than it should otherwise be\cite{bib69}. Such a narrowing of the spectrum may also occur if the detector resolution is not good enough to capture the valleys and peaks in the distribution. The present method of analysis is, therefore has an advantage over the conventional intermittency and multifractality analysis as it does not depend on the detector resolution but rather considers the fractal resolution related to the total energy available\cite{bib10}. \\

\section{Conclusion}
\noindent Experimental data on $^{197}$Au-AgBr collisions are analyzed to study the entropy and multifractal nature of particle production. Since the jets produced may or may not have a uniform spread over the entire azimuth, the two types of events $-$ the ring-like and the jet-like, are identified and analyzed separately to study the entropy production and the degree of multifractality  present in the two categories of events. It is observed that the entropy first grows with increasing width of \et windows and then acquire saturation indicating the presence of the large amount of entropy around mid-rapidity. The decreasing trend of $D_q$ with q supports the multifractal nature of the multiplicity distribution. The value of multifractal specific heat observed in the present study for $^{197}$Au-AgBr collisions is nearly the same as those observed earlier at {\footnotesize AA} collisions at different beam energies. This supports the idea that the parameter may be taken as the universal characteristic of multiparticle production. All these features of data are observed to be nearly reproduced by \urq model. Analysis of ring-like and jet-like events carried out separately reveal that the entropy production in ring-like events is significantly higher than that in jet-like events. The values of multifractal specific heat are also found to be much larger for ring-like events. The smaller values of generalized dimensions, $D_q$ and narrower $f(\alpha_q)$ spectrum observed for jet-like events, in comparison to those for ring-like events, suggest that multifractality is more pronounced in such events which will have \et distributions with pronounced peaks and valleys than those in the case of jet-like events.\\

\end{document}